\documentclass[onecolumn]{pasj00}
\newcommand{\bm}{\boldmath}

\newcommand{\bml}{\mbox{\bm $l$}}

\newcommand{\bmv}{\mbox{\bm $v$}}

\newcommand{\bmnabla}{\mbox{\bm $\nabla$}}
\newcommand{\bmF}{\mbox{\bm $F$}}

\begin{document}
\SetRunningHead{J.\ Fukue}
{Milne-Eddington Solutions for Relativistic Plane-Parallel Flows}
\Received{yyyy/mm/dd}
\Accepted{yyyy/mm/dd}
%tex          2007 1109
%referee      2008 0124
%editing      2007

\title{Milne-Eddington Solutions for Relativistic Plane-Parallel Flows}

%%% begin:list of authors
\author{Jun \textsc{Fukue}} %   \thanks{}}
\affil{Astronomical Institute, Osaka Kyoiku University, 
Asahigaoka, Kashiwara, Osaka 582-8582}
\email{fukue@cc.osaka-kyoiku.ac.jp}

%\author{B-Firstname \textsc{B-Familyname}}
%\affil{B-Address of Institute}\email{bbbbb@xxx.xxx.xx.xx}
%\and
%\author{C-Firstname {\sc C-Familyname}}
%\affil{C-Address of Institute}\email{ccccc@xxx.xxx.xx.xx}
%%% end:list of authors

%% `\KeyWords{}' always has to be placed before `\maketitle'.
\KeyWords{
accretion, accretion disks ---
astrophysical jets ---
%black holes physics ---
%galaxies: active ---
gamma-ray bursts ---
%X-rays: individual (SS~433, GRS~1915$+$105, GRO~J1655$-$40) ---
radiative transfer ---
relativity
%X-rays: stars
} %Do NOT move this preamble from here!

\maketitle

%\newpage

\begin{abstract}
Radiative transfer in a relativistic plane-parallel flow,
e.g., an accretion disk wind, 
is examined in the fully special relativistic treatment.
Under the assumption of a constant flow speed,
for the relativistically moving atmosphere
we analytically obtain generalized Milne-Eddington solutions
of radiative moment equations;
the radiation energy density, the radiative flux, and the radiation pressure.
In the static limit
these solutions reduce to 
the traditional Milne-Eddington ones 
for the plane-parallel static atmosphere,
whereas the source function nearly becomes constant
as the flow speed increases.
Using the analytical solutions,
we analytically integrate the relativistic transfer equation
to obtain the specific intensity.
This specific intensity also reduces to the Milne-Eddinton case
in the static limit,
while the emergent intensity is strongly enhanced
toward the flow direction due to the Doppler and aberration effects
as the flow speed increases (relativistic peaking).
%Ï•ª'µ'ÄŠm"F'µ'½'±'Æ
\end{abstract}

\section{Introduction}

Radiative transfer problems in astrophysics
have been very widely and extensively studied
since 1920's (Milne 1921; Eddington 1926; Kosirev 1934; Chandrasekhar 1934).
There are many excellent monographs
on radiative transfer and radiation hydrodynamics
(e.g., Chandrasekhar 1960; Mihalas 1970; Rybicki, Lightman 1979;
Mihalas, Mihalas 1984; Shu 1991; Peraiah 2002; Castor 2004).
In the early days
the researchers were focused their attention
on radiative transfer in a static atmosphere
of plane-parallel and spherical cases,
relating to stellar atmospheres and nebulae.
Later
they examined radiative transfer in extended and expanding atmospheres,
relating to stellar winds, nova explosions, and so on.

Todays
we know there are many energetic outflows in the universe,
which are blown off from the central luminous sources
with mildly to highly relativistic speeds.
For example, 
accretion disk winds can emanate from cataclysmic variables (CVs),
supersoft X-ray sources (SSXSs), X-ray binaries (XBs),
and active galactic nuclei (AGNs);
relativistic jets are ejected from
microquasars ($\mu$QSOs) and quasars (QSOs);
relativistic outflows occur
in supernovae (SNe), hypernovae (HNe), and gamma-ray bursts (GRBs).

%Furthermore,
%energetic emissions from relativistic jets have been examined,
%relating to, e.g., gamma-ray blazars and gamma-ray bursts
%(Dermer, Schlickeiser 1993, 2002; Dermer 1998;
%B\"ottcher, Dermer 2002; Dermer et al. 2007).

In such circumstances,
radiative transfer problems in relativistically moving media
become more and more important.
However, there are little theoretical and/or analytical studies.
In particular, radiative transfer in the accretion disk wind
has not been well considered.
For example,
transformation properties of disk radiation fields
in the proper frame of a relativistic jet
were examined by, e.g., Dermer and Schlickeiser (2002).
In these earlier works, however,
the radiation fields are set to be external sources,
and the radiation transfer was not considered.
Recently,
radiative transfer in a moving disk atmosphere
was firstly investigated
in the subrelativistic regime
(Fukue 2005a, 2006a),
and in the relativistic regime
(Fukue 2005b, 2006b; Fukue, Akizuki 2006b).
In these studies, however, 
only the radiative moments were obtained
under the moment formalism,
and the specific intensity was not solved.
Hence,
the specific intensity in a plane-parallel moving atmosphere was obtained 
in the subrelativistic regime (Fukue 2007),
and in the relativistic regime (Fukue 2008a).

In the previous study (Fukue 2008a),
under the assumptions that both the flow speed and the source function
are constant, we were able to obtain the analytical solutions.
Since we assumed that the source function is constant, however,
the solutions obtained cannot be compared with the traditional
solutions such as Milne-Eddington ones for a static atmosphere.

Thus,
in this paper
we loosen the assumptions and extend the previous work.
Namely, 
we examine radiative transfer in the relativistic plane-parallel flow,
such as an accretion disk wind,
which is assumed to blow off from the luminous disk
in the vertical direction (plane-parallel approximation),
at a {\it constant} speed.

%In contrast to the static atmosphere,
%in the moving atmosphere
%the boundary condition at the surface of zero optical depth
%should be modified (Fukue 2005a, b).
%Moreover,
%the usual Eddington approximation violates
%in the highly relativistic flow (Fukue 2005b;
%see also 
%Turolla, Nobili 1988; Nobili et al. 1991;
%Turolla et al. 1995; Dullemond 1999),
%and the velocity-dependent variable Eddington factor
%was proposed (Fukue 2006b for a plane-parallel case;
%Akizuki, Fukue 2007 for a spherical case).

%Radiation hydrodynamical (RHD) simulations were also performed
%for radiation-dominated supercritical disks with winds
%by several researchers
%(Eggum et al. 1985, 1988; Okuda et al. 1997, 2005; 
%Okuda, Fujita 2000; Okuda 2002;
%Ohsuga et al. 2005; Ohsuga 2006).
%In these current studies of RHD simulations for disks and winds,
%they were done in the subrelativistic regime
%up to the order of $(v/c)^1$, using the moment formalism
%and the flux-limited diffusion (FLD) approximation
%(Levermore, Pomraning 1981).
%The flux-limited diffusion method provides
%good approximations to the exact solutions
%but only if they are derived from transfer equations
%in which terms of the order of $(v/c)^2$ or higher have been retained
%(Yin, Miller 1995).

In the next section
we describe the basic equations.
In sections 3 and 4, we show analytical solutions
of moment equations, and of transfer equation, respectively.
We then briefly discuss the consistency of the present
analytical solutions in section 5.
The final section is devoted to concluding remarks.

%%%%%%%%%%%%%%%%%%%%%%%%%%%%%%%%%%%%%%%%%%

\section{Relativistic Radiative Transfer Equation
and Moment Equations}

Let us suppose a plane-parallel moving atmosphere,
e.g., a luminous flat disk wind.
The radiation energy is transported in the vertical direction,
and the gas, itself, also moves in the vertical direction
as a {\it plane-parallel outflow}
due to the action of radiation pressure.
For simplicity, in the present paper,
the radiation field is considered to be sufficiently intense that
both the gravitational field of, e.g., the central object
and the gas pressure can be ignored.
We also assume the gray approximation,
where the opacities do not depend on the frequency.
As for the order of the flow velocity $v$,
we consider the fully special relativistic regime.

The radiative transfer equations 
are given in several literatures
(Chandrasekhar 1960; Mihalas 1970; Rybicki, Lightman 1979;
Mihalas, Mihalas 1984; Shu 1991; Kato et al. 1998, 2008;
Mihalas, Auer 2001;
Peraiah 2002; Castor 2004).
The basic equations for relativistic radiation hydrodynamics
are given in, e.g., the appendix E of Kato et al. (2008)
in general and vertical forms,
where the quantities are expressed in the inertial frame.
The relativistic radiation hydrodynamic equations
in the laboratory frame are also derived and discussed
in the seminal paper by Mihalas and Auer (2001) in some details.

In a general form
the radiative transfer equation in the inertial (fixed) frame
is expressed as
\begin{eqnarray}
%\lefteqn{
   \frac{1}{c} \frac{\partial I}{\partial t} + 
   \left( \bml \cdot \bmnabla \right) I
    &=&  \rho \gamma^{-3} \left( 1- \frac{\bmv\cdot\bml}{c} \right)^{-3}
%}
         \left[
            \frac{j_0}{4\pi}
            -\left(\kappa_0^{\rm abs}+\kappa_0^{\rm sca}\right)
             \gamma^{4} \left( 1- \frac{\bmv\cdot\bml}{c} \right)^{4} I
         \right.
\nonumber \\
   &&       
\hspace{2cm}
            + \frac{\kappa_0^{\rm sca} }{4\pi} \frac{3}{4} 
              \gamma^{-2} \left( 1- \frac{\bmv\cdot\bml}{c} \right)^{-2}
         \left\{
         \gamma^4 \left[ \left( 1- \frac{\bmv\cdot\bml}{c} \right)^2
         + \left( \frac{v^2}{c^2} -\frac{\bmv\cdot\bml}{c} \right)^2 \right] cE
%\nonumber \\
         \right.
\nonumber \\
   &&  
\hspace{2cm}
        + 2\gamma^2 \left( \frac{v^2}{c^2} -\frac{\bmv\cdot\bml}{c} \right)
           {\bmF\cdot\bml}
         -2\gamma^4 \left[
          \left( 1- \frac{\bmv\cdot\bml}{c} \right)^2
          + \left( 1- \frac{\bmv\cdot\bml}{c} \right)
            \left( \frac{v^2}{c^2} -\frac{\bmv\cdot\bml}{c} \right) \right]
            \frac{\bmv\cdot\bmF}{c}
\nonumber \\
   &&  
\hspace{2cm}
         \left. \left.
         + l_{i} l_{j} cP^{ij}
         - 2\gamma^2 \left( 1- \frac{\bmv\cdot\bml}{c} \right)
           {v_i l_j P^{ij}}
         + 2\gamma^4 \left( 1- \frac{\bmv\cdot\bml}{c} \right)^2
           \frac{v_i v_j P^{ij}}{c}
         \right\}
         \right],
\label{itransf}
\end{eqnarray}
Here, $\bmv$ is the flow velocity, $c$ is the speed of light, and
$\gamma$ ($=1/\sqrt{1-v^2/c^2}$) is the Lorentz factor.
In the left-hand side
the frequency-integrated specific intensity $I$ and
the direction cosine $\bml$ are quantities
measured in the inertial (fixed) frame.
In the right-hand side, 
the mass density $\rho$,
the frequency-integrated mass emissivity $j_0$, 
the frequency-integrated mass absorption coefficient $\kappa_0^{\rm abs}$,
and
the frequency-integrated mass scattering coefficient $\kappa_0^{\rm sca}$
are quantities measured in the comoving (fluid) frame,
whereas
the frequency-integrated radiation energy density $E$,
the frequency-integrated radiative flux $\bmF$, and
the frequency-integrated radiation stress tensor $P^{ij}$
are quantities measured in the inertial (fixed) frame.

In the plane-parallel geometry with the vertical axis $z$
and the direction cosine $\mu$ ($=\cos \theta$),
the transfer equation is expressed as
\begin{eqnarray}
   \mu \frac{dI}{dz}
    &=&  \rho \frac{1}{\gamma^3 (1-\beta \mu)^3}
         \left[
            \frac{j_0}{4\pi}
            -\left(\kappa_0^{\rm abs}+\kappa_0^{\rm sca}\right)
             \gamma^{4} \left( 1-\beta \mu \right)^{4} I
         \right.
          + \frac{\kappa_0^{\rm sca} }{4\pi} \frac{3}{4} \gamma^2 
         \left\{ 
           \left[ 1+\frac{(\mu-\beta)^2}{(1-\beta \mu)^2}\beta^2
                   +\frac{(1-\beta^2)^2}{(1-\beta \mu)^2}
                    \frac{1-\mu^2}{2} \right] cE
         \right.
\nonumber \\
   &&
\hspace{3cm}
         - \left[ 1+ \frac{(\mu-\beta)^2}{(1-\beta \mu)^2} \right] 2F \beta
         \left. \left.
         + \left[ \beta^2 + \frac{(\mu-\beta)^2}{(1-\beta \mu)^2}
                          - \frac{(1-\beta^2)^2}{(1-\beta \mu)^2}
                            \frac{1-\mu^2}{2} \right] cP
         \right\}
         \right],
\label{itransf_pp}
\end{eqnarray}
where $\beta$ ($=v/c$) is the normalized vertical speed, and
$F$ and $P$ are the vertical component
of the radiative flux and the radiation stress tensor
measured in the inertial frame, respectively.

%For the convenience of readers,
%we shall show the full set of radiation hydrodynamical equations
%under the plane-parallel approximation,
%although we do not use all of them in this paper.

For matter, 
the continuity equation, the equation of motion, and the energy equation
become, respectively,
\begin{eqnarray}
   \rho cu &=& \rho \gamma \beta c = J ~(={\rm const.}),
\label{continuity_pp}
\\
   c^2u\frac{du}{dz} &=& c^2 \gamma^4 \beta \frac{d\beta}{dz}
                      = -\frac{d\psi}{dz} 
               - \gamma^2 \frac{c^2}{\varepsilon + p}\frac{dp}{dz}
              +\frac{\rho c^2}{\varepsilon + p}
                   \frac{\kappa_0^{\rm abs}+\kappa_0^{\rm sca}}{c} \gamma^3 
                    \left[ \frac{}{}
                F (1+\beta^2) - (cE+cP)\beta \frac{}{} \right],
\label{motion_pp}
\\
   0 &=& \frac{q^+}{\rho} - \left( \frac{}{} j_0 - \kappa_0^{\rm abs} cE \gamma^2 
                  - \kappa_0^{\rm abs} cP u^2
                  + 2 \kappa_0^{\rm abs} F \gamma u \frac{}{} \right),
\label{energy_pp}
\end{eqnarray}
where $u$ ($=\gamma \beta$) is the vertical four velocity, 
$J$ the mass-loss rate per unit area,
$\psi$ the gravitational potential, 
$\varepsilon$ the internal energy per unit proper volume,
$p$ the gas pressure, and
$q^+$ the internal heating.
In the energy equation (\ref{energy_pp})
the advection terms in the left-hand side
are dropped under the present cold approximation
(radiative equilibrium).
We do not use the equation of motion
since we will assume that the flow speed is constant.

For radiation,
the zeroth and first moment equations,
and the closure relation become, respectively,
\begin{eqnarray}
   \frac{dF}{dz} &=& \rho \gamma
         \left[ \frac{}{} j_0 - \kappa_0^{\rm abs} cE
                 + \kappa_0^{\rm sca} (cE+cP) \gamma^2 \beta^2  
        + \kappa_0^{\rm abs} F \beta
               - \kappa_0^{\rm sca} F ( 1+ \beta^2 )\gamma^2 \beta \frac{}{} \right],
\label{mom0_pp}
\\
   \frac{dP}{dz} &=& \frac{\rho \gamma}{c} 
         \left[ \frac{}{} j_0 \beta - \kappa_0^{\rm abs}  F
                  + \kappa_0^{\rm abs} cP \beta 
      - \kappa_0^{\rm sca} F \gamma^2 (1+\beta^2)
               + \kappa_0^{\rm sca} (cE+cP)\gamma^2 \beta \frac{}{} \right],
\label{mom1_pp}
\\
   cP(1-f\beta^2) &=& cE(f-\beta^2) + 2F\beta(1-f),
\label{closure_pp}
\end{eqnarray}
where 
$f(\tau, \beta)$ is the variable Eddington factor,
which is defined by $f=P_{\rm co}/E_{\rm co}$,
$E_{\rm co}$ and $P_{\rm co}$ being the comoving quantities,
and generally depends on the velocity 
and its gradient as well as the optical depth (Fukue 2006b, 2008b).
In the present approximation, however, $f$ is automatically determined
to be 1/3 as will be shown later.

Eliminating $j_0$ using the energy equation (\ref{energy_pp}),
the transfer equation (\ref{itransf_pp}) becomes
\begin{eqnarray}
   \mu \frac{dI}{dz}
    &=&  \rho \frac{1}{\gamma^3 (1-\beta \mu)^3}
         \left[
            -\left(\kappa_0^{\rm abs}+\kappa_0^{\rm sca}\right)
             \gamma^{4} \left( 1-\beta \mu \right)^{4} I
          +  \frac{q^+}{4\pi \rho}
          + \frac{\kappa_0^{\rm abs} }{4\pi} \gamma^2 
            (cE - 2F\beta + \beta^2 cP)
         \right.
\nonumber \\
    &&
\hspace{3cm}
          + \frac{\kappa_0^{\rm sca} }{4\pi} \frac{3}{4} \gamma^2 
         \left\{ 
           \left[ 1+\frac{(\mu-\beta)^2}{(1-\beta \mu)^2}\beta^2
                   +\frac{(1-\beta^2)^2}{(1-\beta \mu)^2}
                    \frac{1-\mu^2}{2} \right] cE
         \right.
\nonumber \\
   &&
\hspace{5cm}
         - \left[ 1+ \frac{(\mu-\beta)^2}{(1-\beta \mu)^2} \right] 2F \beta
         \left. \left.
         + \left[ \beta^2 + \frac{(\mu-\beta)^2}{(1-\beta \mu)^2}
                          - \frac{(1-\beta^2)^2}{(1-\beta \mu)^2}
                            \frac{1-\mu^2}{2} \right] cP
         \right\}
         \right].
\label{itransf_pp2}
\end{eqnarray}

Introducing the optical depth defined by
\begin{equation}
   d\tau = - \left(\kappa_0^{\rm abs}+\kappa_0^{\rm sca}\right) \rho dz,
\end{equation}
the transfer equation (\ref{itransf_pp2}) finally becomes
\begin{eqnarray}
   \mu \frac{dI}{d\tau}
    &=&  \frac{1}{\gamma^3 (1-\beta \mu)^3}
         \left[
             \gamma^{4} \left( 1-\beta \mu \right)^{4} I
          - \frac{1}{4\pi}
            \frac{\kappa_0^{\rm abs} }{\kappa_0^{\rm abs}+\kappa_0^{\rm sca}}
            \gamma^2 
            (cE - 2F\beta + \beta^2 cP)
         \right.
\nonumber \\
    &&
\hspace{1.8cm}
          - \frac{1}{4\pi}
         \frac{\kappa_0^{\rm sca} }{\kappa_0^{\rm abs}+\kappa_0^{\rm sca}}
         \frac{3}{4} \gamma^2 
         \left\{ 
           \left[ 1+\frac{(\mu-\beta)^2}{(1-\beta \mu)^2}\beta^2
                   +\frac{(1-\beta^2)^2}{(1-\beta \mu)^2}
                    \frac{1-\mu^2}{2} \right] cE
         \right.
\nonumber \\
   &&
\hspace{4.5cm}
         - \left[ 1+ \frac{(\mu-\beta)^2}{(1-\beta \mu)^2} \right] 2F \beta
         \left. \left.
         + \left[ \beta^2 + \frac{(\mu-\beta)^2}{(1-\beta \mu)^2}
                          - \frac{(1-\beta^2)^2}{(1-\beta \mu)^2}
                            \frac{1-\mu^2}{2} \right] cP
         \right\}
         \right]
\nonumber \\
    &=&   \gamma \left( 1-\beta \mu \right) I
          - \frac{1-A}{4\pi}
          \frac{1}{\gamma (1-\beta \mu)^3}
            (cE - 2F\beta + \beta^2 cP)
\nonumber \\
    &&
\hspace{1.8cm}
          - \frac{A}{4\pi} \frac{3}{4}
          \frac{1}{\gamma (1-\beta \mu)^3}
         \left\{ 
           \left[ 1+\frac{(\mu-\beta)^2}{(1-\beta \mu)^2}\beta^2
                   +\frac{(1-\beta^2)^2}{(1-\beta \mu)^2}
                    \frac{1-\mu^2}{2} \right] cE
         \right.
\nonumber \\
   &&
\hspace{4.5cm}
         - \left[ 1+ \frac{(\mu-\beta)^2}{(1-\beta \mu)^2} \right] 2F \beta
         \left.
         + \left[ \beta^2 + \frac{(\mu-\beta)^2}{(1-\beta \mu)^2}
                          - \frac{(1-\beta^2)^2}{(1-\beta \mu)^2}
                            \frac{1-\mu^2}{2} \right] cP
         \right\},
\label{itransf_pp3}
\end{eqnarray}
where
\begin{equation}
   A \equiv
           \frac{\kappa_0^{\rm sca} }{\kappa_0^{\rm abs}+\kappa_0^{\rm sca}}
\end{equation}
is the scattering albedo,
and the internal heating $q^+$ is dropped.

Similarly, moment equations 
(\ref{mom0_pp}) and (\ref{mom1_pp}),
with the help of continuity equation (\ref{continuity_pp})
and the closure relation (\ref{closure_pp}),
become (cf. Fukue 2005b, 2006b)
\begin{eqnarray}
   \frac{dF}{d\tau}
      &=&  \gamma^3 \beta [ F(1+\beta^2) - (cE+cP) \beta]
       =   \gamma \beta \frac{F(f+\beta^2) -cP(1+f)\beta}{f-\beta^2},
\label{mom0_pp3}
\\
   c\frac{dP}{d\tau}
      &=&  \gamma^3 [ F(1+\beta^2) - (cE+cP) \beta]
       =   \gamma \frac{F(f+\beta^2) -cP(1+f)\beta}{f-\beta^2}.
\label{mom1_pp3}
\end{eqnarray}
Here, we dropped the gravitational and pressure forces
as well as the internal heating.

\section{Solutions of Moment Equations}

In the present paper
we assume that the flow speed is constant,
but the source function is not,
and seek the analytical solutions
of moment equations (\ref{mom0_pp3}) and (\ref{mom1_pp3}).
We also assume that the Eddington factor does not depend
on the optical depth.

\subsection{General Solutions}

If equation (\ref{mom0_pp3}) is divided by equation (\ref{mom1_pp3}),
we have
\begin{equation}
   \frac{dF}{cdP} = \beta,
\end{equation}
which is integrated to give
\begin{equation}
   F = \beta cP + F_1 (\beta),
\label{FP_relation}
\end{equation}
where $F_1(\beta)$ is an arbitrary function of $\beta$.

Inserting this relation (\ref{FP_relation})
into equations (\ref{mom0_pp3}) and (\ref{mom1_pp3}),
we can easily integrate equations (\ref{mom0_pp3}) and (\ref{mom1_pp3})
to give the following forms:
\begin{eqnarray}
   F &=& \gamma^2 (1+f) F_1(\beta)
         - C_{\rm F}(\beta) e^{-\frac{\displaystyle \beta}
             {\displaystyle \gamma(f-\beta^2)}\displaystyle \tau},
\label{F1}
\\
   cP &=& \frac{\gamma^2(f+\beta^2)}{\beta} F_1(\beta)
         - C_{\rm P}(\beta) e^{-\frac{\displaystyle \beta}
             {\displaystyle \gamma(f-\beta^2)}\displaystyle \tau},
\label{P1}
\end{eqnarray}
where $C_{\rm F}(\beta)$ and $C_{\rm P}(\beta)$ 
are arbitrary functions of $\beta$.

In order to both solutions (\ref{F1}) and (\ref{P1})
are consistent with each other,
$C_{\rm F}=C_{\rm P}\beta$ and we have
\begin{equation}
   cP = \frac{1}{\beta} \left[ \gamma^2(f+\beta^2) F_1(\beta)
         - C_{\rm F}(\beta) e^{-\frac{\displaystyle \beta}
             {\displaystyle \gamma(f-\beta^2)}\displaystyle \tau} \right],
\label{P2}
\end{equation}
and furthermore, from the closure relation (\ref{closure_pp}),
we also have
\begin{equation}
   cE = \frac{1}{\beta (f-\beta^2)} 
          \left[ (f-\beta^2) (1+f\beta^2) \gamma^2 F_1(\beta)
         - (1-2\beta+f\beta^2) C_{\rm F}(\beta) e^{-\frac{\displaystyle \beta}
             {\displaystyle \gamma(f-\beta^2)}\displaystyle \tau} \right].
\label{E2}
\end{equation}

With the help of arbitrariness in $C_{\rm F}$,
we can express general solutions (\ref{F1}), (\ref{P2}), and (\ref{E2}) as
\begin{eqnarray}
   cE &=& \frac{\gamma^2 F_1}{\beta} 
          \left[ (1+f\beta^2) 
         - (1-2\beta+f\beta^2) C_1(\beta) e^{-\frac{\displaystyle \beta}
             {\displaystyle \gamma(f-\beta^2)}\displaystyle \tau} \right],
\label{E3}
\\
   F  &=& \gamma^2 F_1
          \left[ (1+f) 
         - (f-\beta^2) C_1(\beta) e^{-\frac{\displaystyle \beta}
             {\displaystyle \gamma(f-\beta^2)}\displaystyle \tau} \right],
\label{F3}
\\
   cP &=& \frac{\gamma^2 F_1}{\beta} 
          \left[ (f+\beta^2) 
         - (f-\beta^2) C_1(\beta) e^{-\frac{\displaystyle \beta}
             {\displaystyle \gamma(f-\beta^2)}\displaystyle \tau} \right],
\label{P3}
\end{eqnarray}
where $C_1(\beta)=\gamma^2(f-\beta^2)F_1C_{\rm F}$
is an arbitrary function of $\beta$.

These equations (\ref{E3})--(\ref{P3}) are general solutions
of relativistic moment equations
for plane-parallel vertical flows at a constant speed.
In the limit of $\beta \rightarrow 0$,
these solutions approach the Milne-Eddington ones,
as long as $1-C_{\rm F} \sim \beta$ for small $\beta$.

\subsection{Special Solutions}

Next, we derive more special solutions
under the appropriate boundary conditions.
For the present plane-parallel flow
at a constant flow speed,
the boundary conditions should be imposed
at the flow top, where the optical depth is zero.

At the flow top of a moving photosphere at a relativistic speed,
the usual boundary conditions for a static atmosphere is inadequate,
as already pointed out in Fukue (2005b).
Namely,
the radiation field just above the flow top changes
when the gas itself does move upward,
since the direction and intensity of radiation
change due to the relativistic aberration and Doppler effect
(cf. Kato et al. 1998, 2008; Fukue 2000).
If a flat infinite plane with surface intensity $I_{\rm s}$
in the comoving frame is not static,
but moving upward at a speed $v_{\rm s}$ 
($=c\beta_{\rm s}$, and
the corresponding Lorentz factor is $\gamma_{\rm s}$),
where the subscript s denotes the values at the surface,
then, just above the surface,
the radiation energy density $E_{\rm s}$, 
the radiative flux $F_{\rm s}$, and
the radiation pressure $P_{\rm s}$ measured in the inertial frame
become, respectively,
\begin{eqnarray}
   cE_{\rm s} 
   &=& {2\pi I_{\rm s}\gamma_{\rm s}^2}
       \frac{3+3\beta_{\rm s}+\beta_{\rm s}^2}{3},
\label{Es2}
\\
   F_{\rm s}
   &=& {2\pi I_{\rm s}\gamma_{\rm s}^2}
       \frac{3+8\beta_{\rm s}+3\beta_{\rm s}^2}{6},
\label{Fs2}
\\
   cP_{\rm s}
   &=& {2\pi I_{\rm s}\gamma_{\rm s}^2}
       \frac{1+3\beta_{\rm s}+3\beta_{\rm s}^2}{3}.
\label{Ps2}
\end{eqnarray}
It should be noted that
these boundary conditions imply that
the Eddington factor $f$ defined in the comoving frame is 1/3.

Using these boundary conditions (\ref{Es2})--(\ref{Ps2})
to general solusions (\ref{E3})--(\ref{P3}),
we can determine the arbitrary functions $F_1$ and $C_1$ as
\begin{eqnarray}
   F_1 &=& \pi I_{\rm s} (1+2\beta),
\\
   C_1 &=& \frac{1}{1+2\beta},
\end{eqnarray}
and the Eddington factor $f$ is automatically fixed as 1/3.

Finally, the solutions of moment equations
for relativistic plane-parallel flows at a constant speed
are explicitly written as
\begin{eqnarray}
   \frac{cE}{\pi I_{\rm s}} &=& \frac{\gamma^2}{\beta} 
          \left[ (1+f\beta^2)(1+2\beta) 
         - (1-2\beta+f\beta^2) e^{-\frac{\displaystyle \beta}
             {\displaystyle \gamma(f-\beta^2)}\displaystyle \tau} \right],
\label{E_sol}
\\
   \frac{F}{\pi I_{\rm s}}  &=& \gamma^2
          \left[ (1+f)(1+2\beta) 
         - (f-\beta^2) e^{-\frac{\displaystyle \beta}
             {\displaystyle \gamma(f-\beta^2)}\displaystyle \tau} \right],
\label{F_sol}
\\
   \frac{cP}{\pi I_{\rm s}} &=& \frac{\gamma^2}{\beta} 
          \left[ (f+\beta^2)(1+2\beta) 
         - (f-\beta^2) e^{-\frac{\displaystyle \beta}
             {\displaystyle \gamma(f-\beta^2)}\displaystyle \tau} \right],
\label{P_sol}
\end{eqnarray}
where $f=1/3$.

In the limit of small $\beta$,
these solutions approach
\begin{eqnarray}
   \frac{cE}{\pi I_{\rm s}} &\sim& 2+\frac{1}{f}\tau,
\label{E_sol_beta0}
\\
   \frac{F}{\pi I_{\rm s}}  &\sim& 1+2(1+f)\beta+\beta\tau,
\label{F_sol_beta0}
\\
   \frac{cP}{\pi I_{\rm s}} &\sim& 2f+\tau,
\label{P_sol_beta0}
\end{eqnarray}
which are just the Milne-Eddington solutions
for a static plane-parallel atmosphere.
Hence, the present solutions (\ref{E_sol})--(\ref{P_sol})
are the Milne-Eddington ones
extended to the case for the relativistic plane-parallel flow.

By the Lorentz transformation,
it is easy to obtain the solutions in the comoving frame.
Namely, the solutions of moment equations
in the comoving frame
are explicitly written as
\begin{eqnarray}
   \frac{cE_{\rm co}}{\pi I_{\rm s}} &=& \frac{1}{\pi I_{\rm s}}
         \gamma^2 \left( cE-2\beta F +\beta^2 cP \right)
         = \frac{1}{\beta} \left[ (1+2\beta) 
         - e^{-\frac{\displaystyle \beta}
             {\displaystyle \gamma(f-\beta^2)}\displaystyle \tau} \right],
\label{Eco_sol}
\\
   \frac{F_{\rm co}}{\pi I_{\rm s}}  &=& \frac{1}{\pi I_{\rm s}}
         \gamma^2 \left[ (1+\beta^2) F -\beta (cE+cP) \right]
         = e^{-\frac{\displaystyle \beta}
             {\displaystyle \gamma(f-\beta^2)}\displaystyle \tau},
\label{Fco_sol}
\\
   \frac{cP_{\rm co}}{\pi I_{\rm s}} &=& \frac{1}{\pi I_{\rm s}}
         \gamma^2 \left( \beta^2 cE - 2\beta F + cP \right)
         = \frac{f}{\beta} \left[ (1+2\beta) 
         - e^{-\frac{\displaystyle \beta}
             {\displaystyle \gamma(f-\beta^2)}\displaystyle \tau} \right].
\label{Pco_sol}
\end{eqnarray}
where $f=1/3$.
Hence, as already stated, in the comoving frame
$P_{\rm co}/E_{\rm co}=f=1/3$, exactly.

\begin{figure}
  \begin{center}
  \FigureFile(80mm,80mm){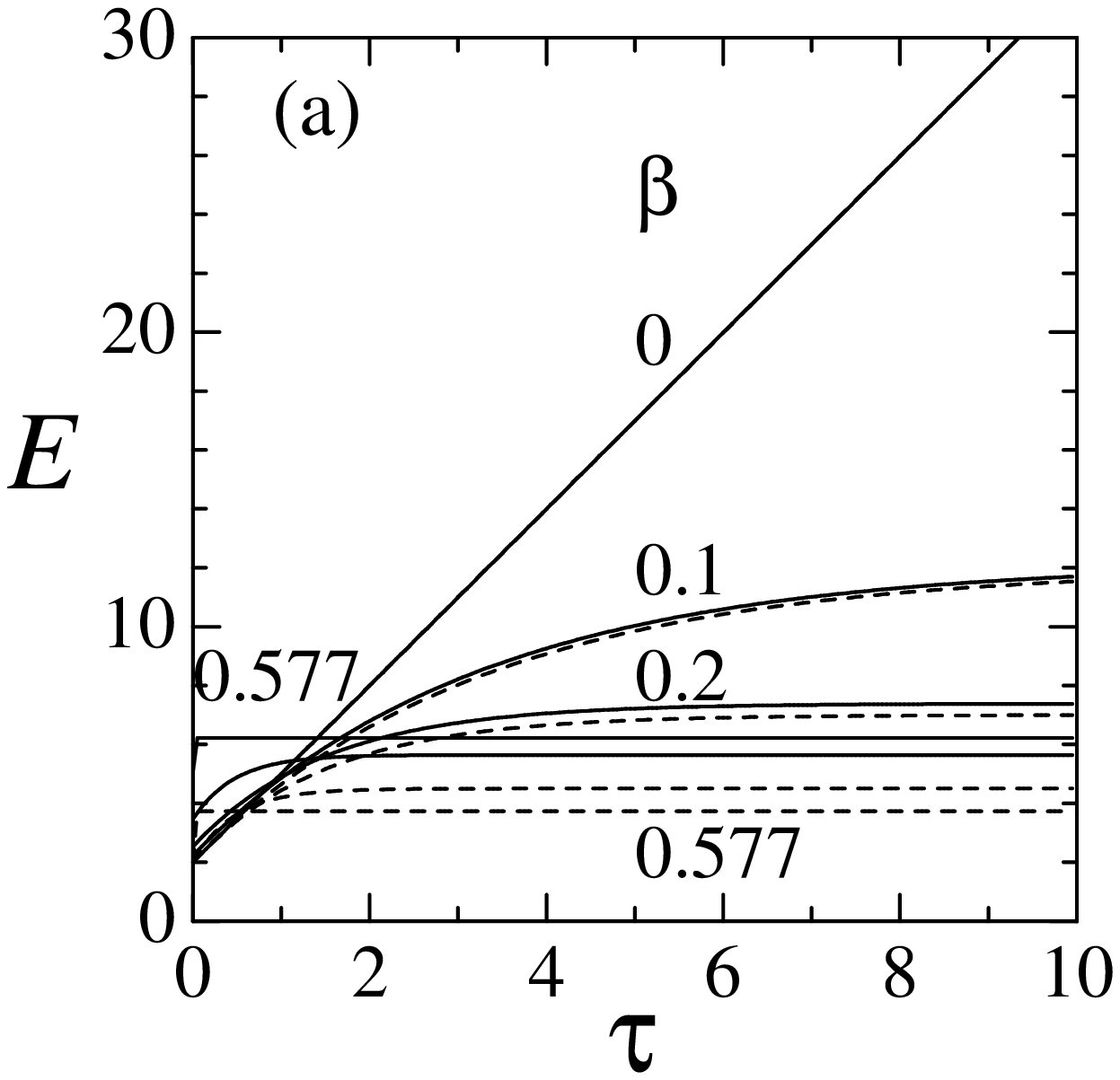}
  \FigureFile(80mm,80mm){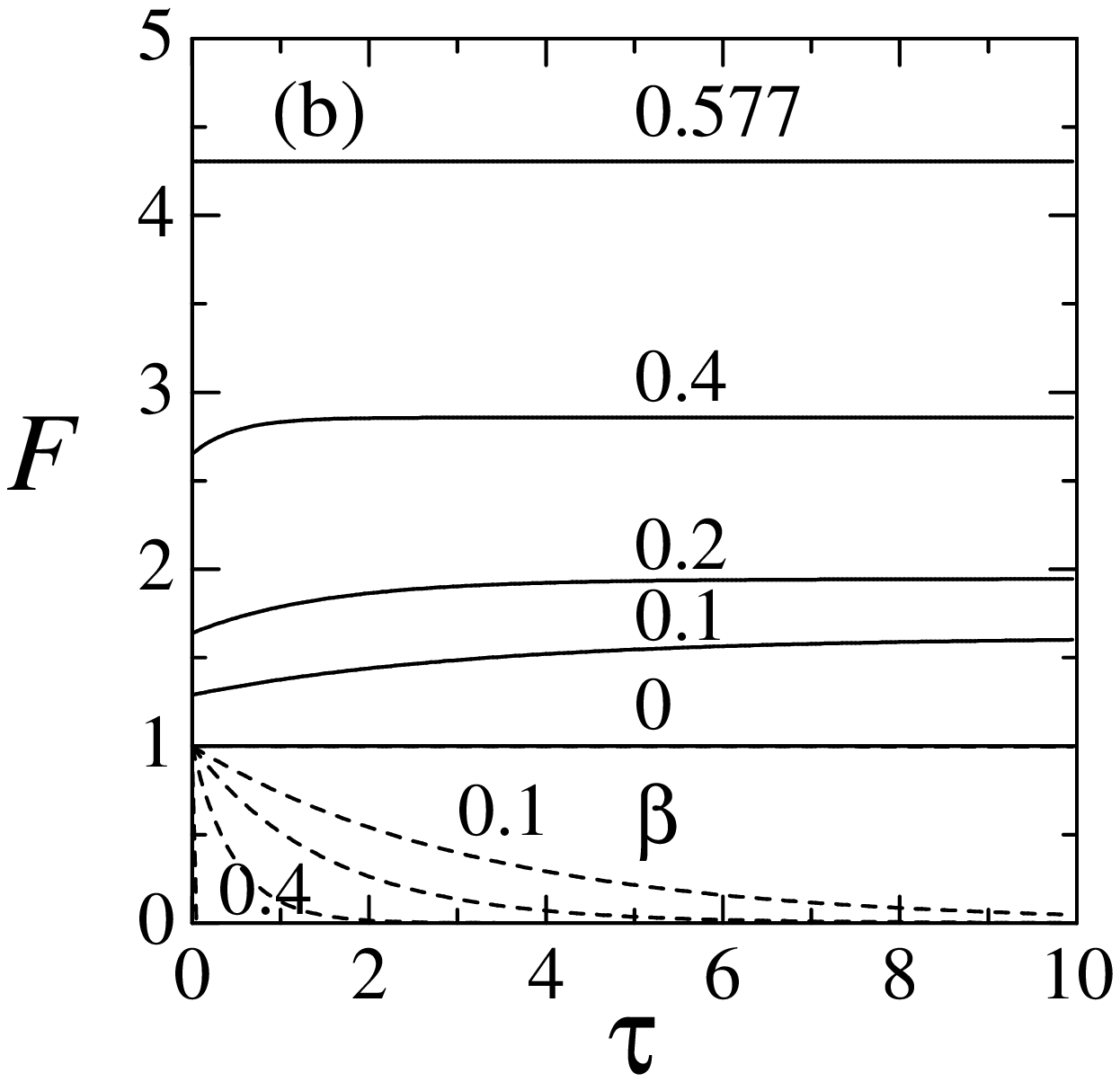}
  \FigureFile(80mm,80mm){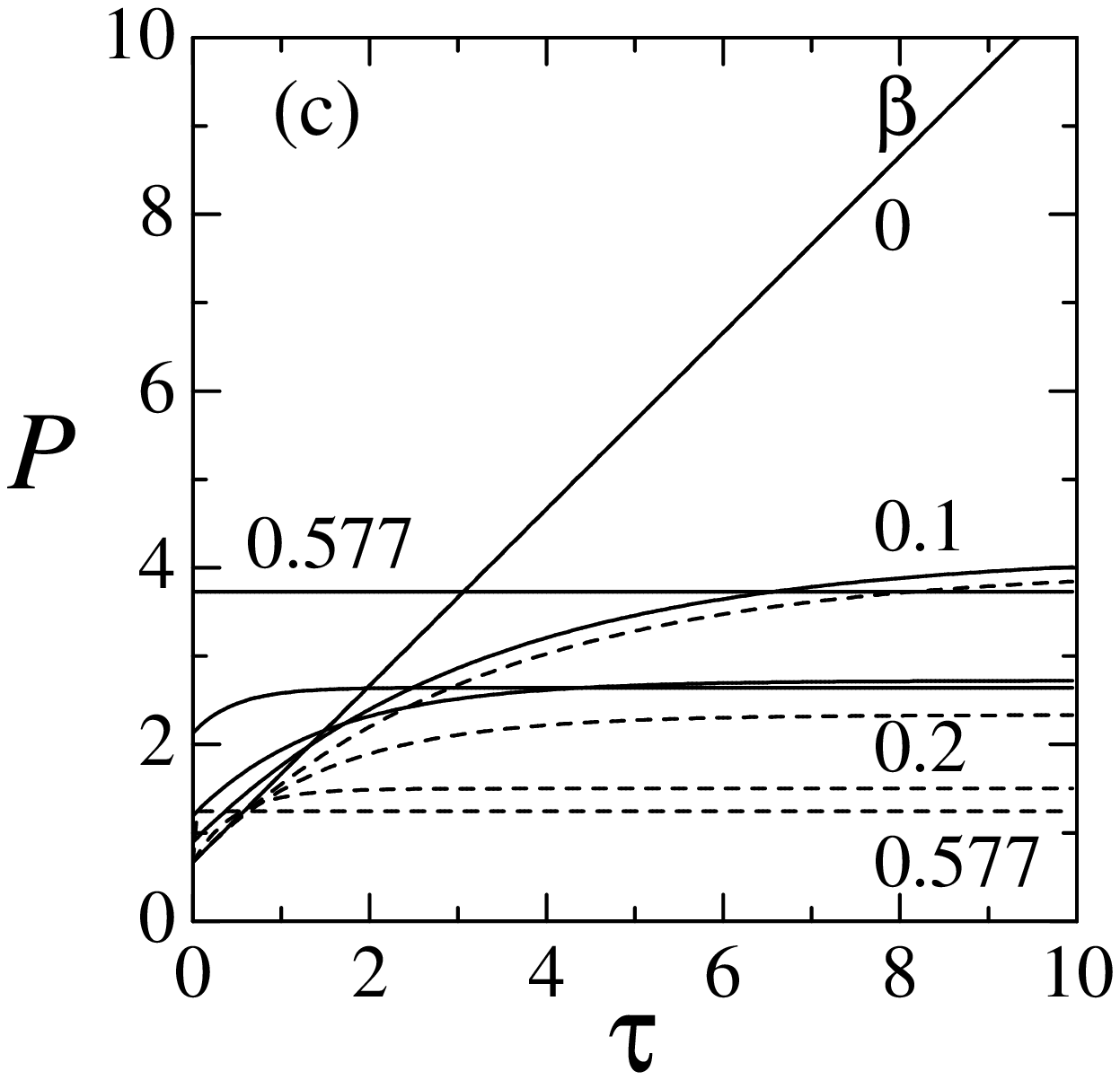}
  \end{center}
\caption{
Milne-Eddington solutions for relativistic plane-parallel flows:
(a) Normalized radiation energy density,
(b) normalized radiative flux, and
(c) normalized radiation pressure.
The solid curves represent the quantities in the inertial frame,
whereas the dashed ones mean those in the comoving frame.
The values of $\beta$ are
0, 0.1, 0.2, 0.4, and 0.577.
}
\end{figure}

In figure 1,
Milne-Eddington solutions of moment equations
for relativistic plane-parallel flows
are shown as a function of the optical depth $\tau$:
(a) Radiation energy density $E$,
(b) radiative flux $F$, and
(c) radiation pressure $P$,
where the radiation quantities are normalized by $\pi I_{\rm s}$.
The solid curves represent the quantities in the inertial frame,
whereas the dashed ones mean those in the comoving frame.
The flow speed $\beta$ are set to be
0, 0.1, 0.2, 0.4, and 0.577.

In figure 1a,
the radiation energy density $E$ normalized by $\pi I_{\rm s}/c$
is plotted for various $\beta$.
As was already stated, and as is easily seen in figure 1a,
when the flow speed is small,
the solution reduces to the usual Milne-Eddington one
for a static plane-parallel atmosphere;
$cE/(\pi I_{\rm s}) \sim 2+3\tau$.
As the flow speed increases,
the radiation energy density becomes constant.
This means that the radiation energy itself is advected
along with the flow, and it is constant without the internal heating.
Furthermore, comparing the radiation energy density
in the inertial frame (solid curves) and those
in the comoving frame (dashed ones),
the radiation energy density in the inertial frame
is enhanced due to the relativistic motion.

In figure 1b,
the radiative flux $F$ normalized by $\pi I_{\rm s}$
is plotted for various $\beta$.
When the flow speed is small,
the solution reduces to the usual Milne-Eddington one
for a static plane-parallel atmosphere;
$F/(\pi I_{\rm s}) \sim 1$.
As the flow speed increases,
the radiative flux in the inertial frame (solid curves)
is enhanced more and more
due to the relativistic effect.
The radiative flux in the comoving frame (dashed ones),
on the other hand,
reduces exponentially.
This means that in the relativistically moving plane-parallel flows
without the internal heating
the radiative flux in the comoving frame vanishes
since the radiation energy (and pressure) is advected with the flow
and there is no gradient in the radiation pressure.

In figure 1c,
the radiation pressure $P$ normalized by $\pi I_{\rm s}/c$
is plotted for various $\beta$.
When the flow speed is small,
the solution reduces to the usual Milne-Eddington one
for a static plane-parallel atmosphere;
$cP/(\pi I_{\rm s}) \sim 2/3+\tau$.
As the flow speed increases,
the radiation pressure becomes constant.
Furthermore, comparing the radiation energy pressure
in the inertial frame (solid curves) and those
in the comoving frame (dashed ones),
the radiation pressure in the inertial frame
is enhanced due to the relativistic motion.

It should be noted that
in the present solution
the Eddington factor $f$ is set to be 1/3,
and therefore,
the flow speed $\beta$ is restricted within
$\beta < 1/\sqrt{3} \sim 0.5774$
(cf. Fukue 2005a).

\section{Solutions of Transfer Equation}

Since we obtain the analytical forms 
of radiation moments, and we assume
that the flow speed is constant,
we can analytically integrate
the relativistic radiation transfer equation (\ref{itransf_pp3}).

By inserting the solutions (\ref{E_sol})--(\ref{P_sol})
into the transfer equation (\ref{itransf_pp3}),
and setting $f=1/3$,
the transfer equation (\ref{itransf_pp3}) is rewritten as
\begin{eqnarray}
   \mu \frac{dI}{d\tau}
    &=&   \gamma \left( 1-\beta \mu \right) I
          - \frac{\pi I_{\rm s}}{4\pi}
          \frac{1}{\gamma^3 (1-\beta \mu)^3}
          \frac{1}{\beta} \left[ (1+2\beta) 
         - e^{-\frac{\displaystyle \beta}
             {\displaystyle \gamma(f-\beta^2)}\displaystyle \tau} \right],
\label{itransf_pp4}
\end{eqnarray}
where $f=1/3$.

It is straightforward to integrate this equation (\ref{itransf_pp4})
(cf. Fukue, Akizuki 2006b; Fukue 2007).
After several partial integrations,
we obtain both an outward intensity $I(\tau, \mu, \beta)$ ($\mu>0$),
\begin{eqnarray}
   I(\tau, \mu, \beta)
   &=& \frac{\pi I_{\rm s}}{4\pi} \frac{1}{\gamma^4 (1-\beta \mu)^4}
       \frac{1}{\beta} \left[ (1+2\beta) 
         - \frac{1}{1+\frac{\displaystyle \beta\mu}
                    {\displaystyle \gamma^2(1-\beta\mu)(f-\beta^2)}}
           e^{-\frac{\displaystyle \beta}
                    {\displaystyle \gamma(f-\beta^2)}
              \displaystyle \tau} \right]
\nonumber \\
   && -\frac{\pi I_{\rm s}}{4\pi} \frac{1}{\gamma^4 (1-\beta \mu)^4}
       \frac{1}{\beta} \left[ (1+2\beta) 
         - \frac{1}{1+\frac{\displaystyle \beta\mu}
                    {\displaystyle \gamma^2(1-\beta\mu)(f-\beta^2)}}
           e^{-\frac{\displaystyle \beta}
                    {\displaystyle \gamma(f-\beta^2)}
              \displaystyle \tau_0} \right]
           e^{\frac{\displaystyle \gamma(1-\beta\mu)}
                   {\displaystyle \mu}
              \displaystyle (\tau - \tau_0)}
\nonumber \\
   && + I(\tau_0, \mu)
           e^{\frac{\displaystyle \gamma(1-\beta\mu)}
                   {\displaystyle \mu}
              \displaystyle (\tau - \tau_0)}
\nonumber \\
   &\sim& \frac{\pi I_{\rm s}}{4\pi} \frac{1}{\gamma^4 (1-\beta \mu)^4}
       \frac{1}{\beta} \left[ (1+2\beta) 
         - \frac{1}{1+\frac{\displaystyle \beta\mu}
                    {\displaystyle \gamma^2(1-\beta\mu)(f-\beta^2)}}
           e^{-\frac{\displaystyle \beta}
                    {\displaystyle \gamma(f-\beta^2)}
              \displaystyle \tau} \right]
       ~~~~~{\rm for~large~}\tau_0,
\label{i_plus}
\end{eqnarray}
and an inward intensity $I(\tau, \mu, \beta)$ ($\mu<0$),
\begin{eqnarray}
   I(\tau, \mu, \beta)
   &=& \frac{\pi I_{\rm s}}{4\pi} \frac{1}{\gamma^4 (1-\beta \mu)^4}
       \frac{1}{\beta} \left[ (1+2\beta) 
         - \frac{1}{1+\frac{\displaystyle \beta\mu}
                    {\displaystyle \gamma^2(1-\beta\mu)(f-\beta^2)}}
           e^{-\frac{\displaystyle \beta}
                    {\displaystyle \gamma(f-\beta^2)}
              \displaystyle \tau} \right]
\nonumber \\
   && -\frac{\pi I_{\rm s}}{4\pi} \frac{1}{\gamma^4 (1-\beta \mu)^4}
       \frac{1}{\beta} \left[ (1+2\beta) 
         - \frac{1}{1+\frac{\displaystyle \beta\mu}
                    {\displaystyle \gamma^2(1-\beta\mu)(f-\beta^2)}}
            \right]
           e^{\frac{\displaystyle \gamma(1-\beta\mu)}
                   {\displaystyle \mu}
              \displaystyle \tau},
\label{i_minus}
\end{eqnarray}
where $I(\tau_0, \mu)$ is the boundary value
at the flow base of the optical depth $\tau_0$.

In the limit of small $\beta$,
this solution reduces to the usual Milne-Eddington one;
e.g., for $\mu>0$
\begin{equation}
   I(\tau, \mu, 0) \sim 
      \frac{\pi I_{\rm s}}{4\pi}
      \left[
         (2+3\tau+3\mu) - (2+3\tau_0+3\mu)
             e^{\frac{\displaystyle \tau-\tau_0}
                     {\displaystyle \mu}}
      \right]
      + I(\tau_0, \mu)
             e^{\frac{\displaystyle \tau-\tau_0}
                     {\displaystyle \mu}}.
\end{equation}

Finally, the emergent intensity $I(0, \mu, \beta)$ ($\mu>0$)
at the flow surface of $\tau=0$ becomes
\begin{eqnarray}
   I(0, \mu, \beta)
   &=& \frac{\pi I_{\rm s}}{4\pi} \frac{1}{\gamma^4 (1-\beta \mu)^4}
       \frac{2 +
                 \frac{\displaystyle (1+2\beta)\mu}
                      {\displaystyle \gamma^2(1-\beta\mu)(f-\beta^2)}}
            {1 +
                 \frac{\displaystyle \beta\mu}
                      {\displaystyle \gamma^2(1-\beta\mu)(f-\beta^2)}}
\nonumber \\
   && -\frac{\pi I_{\rm s}}{4\pi} \frac{1}{\gamma^4 (1-\beta \mu)^4}
       \frac{1}{\beta} \left[ (1+2\beta) 
         - \frac{1}{1+\frac{\displaystyle \beta\mu}
                    {\displaystyle \gamma^2(1-\beta\mu)(f-\beta^2)}}
           e^{-\frac{\displaystyle \beta}
                    {\displaystyle \gamma(f-\beta^2)}
              \displaystyle \tau_0} \right]
           e^{-\frac{\displaystyle \gamma(1-\beta\mu)}
                   {\displaystyle \mu}
              \displaystyle \tau_0}
\nonumber \\
   && + I(\tau_0, \mu)
           e^{-\frac{\displaystyle \gamma(1-\beta\mu)}
                   {\displaystyle \mu}
              \displaystyle \tau_0}
\nonumber \\
   &\sim& \frac{\pi I_{\rm s}}{4\pi} \frac{1}{\gamma^4 (1-\beta \mu)^4}
       \frac{2 +
                 \frac{\displaystyle (1+2\beta)\mu}
                      {\displaystyle \gamma^2(1-\beta\mu)(f-\beta^2)}}
            {1 +
                 \frac{\displaystyle \beta\mu}
                      {\displaystyle \gamma^2(1-\beta\mu)(f-\beta^2)}}
       ~~~~~{\rm for~large~}\tau_0.
\label{i_emergent}
\end{eqnarray}

\begin{figure}
  \begin{center}
  \FigureFile(80mm,80mm){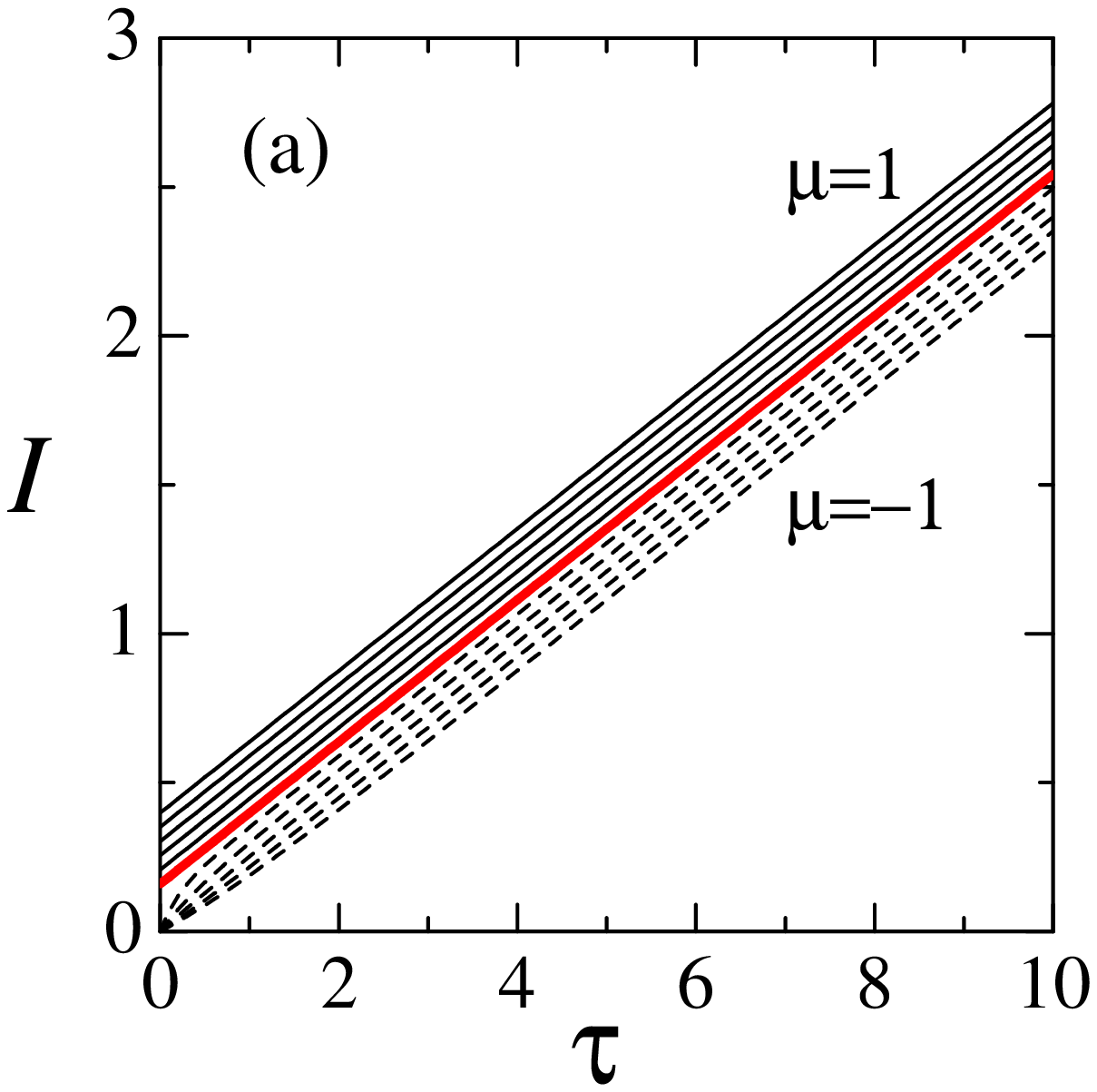}
  \FigureFile(80mm,80mm){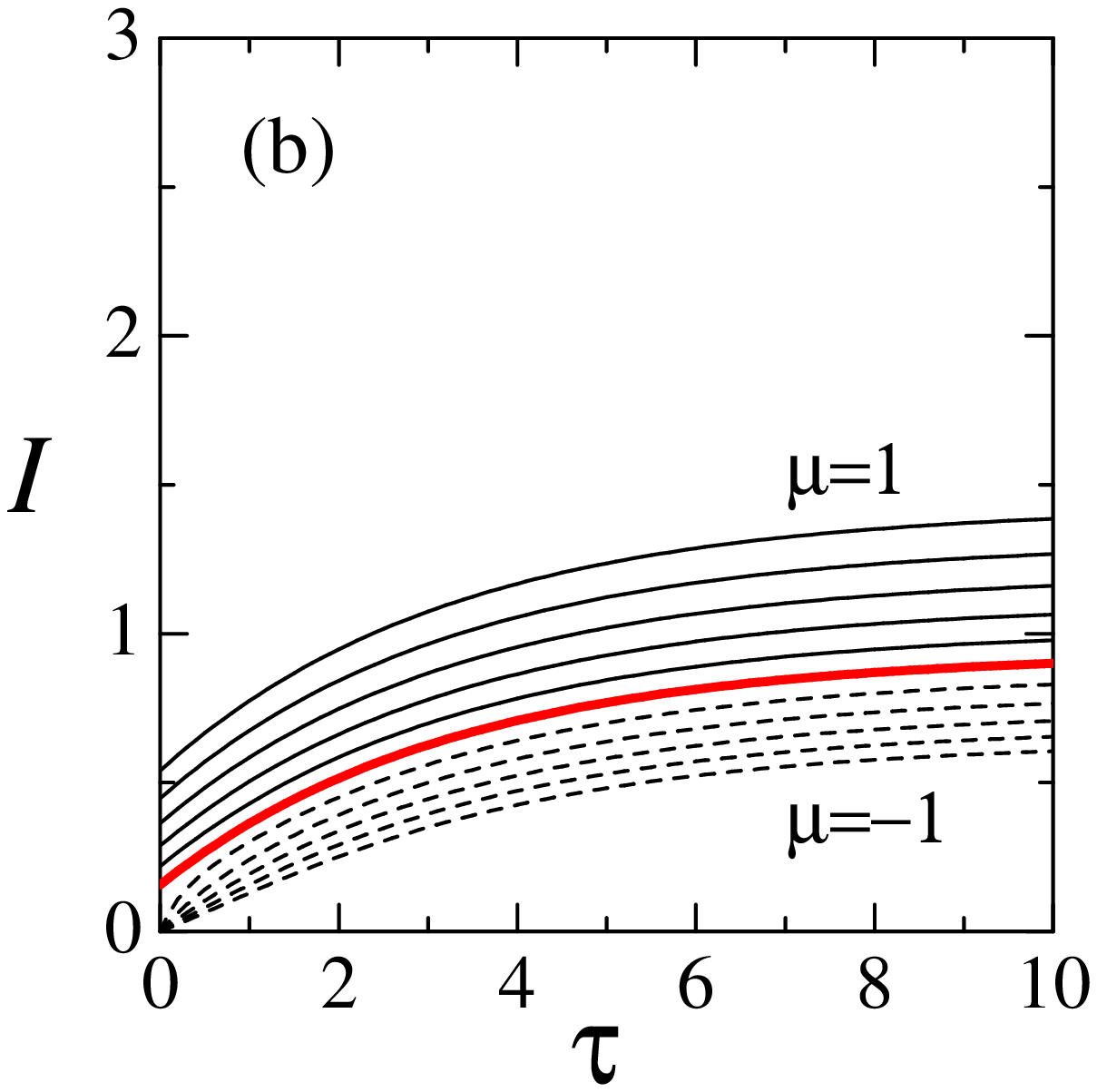}
  \FigureFile(80mm,80mm){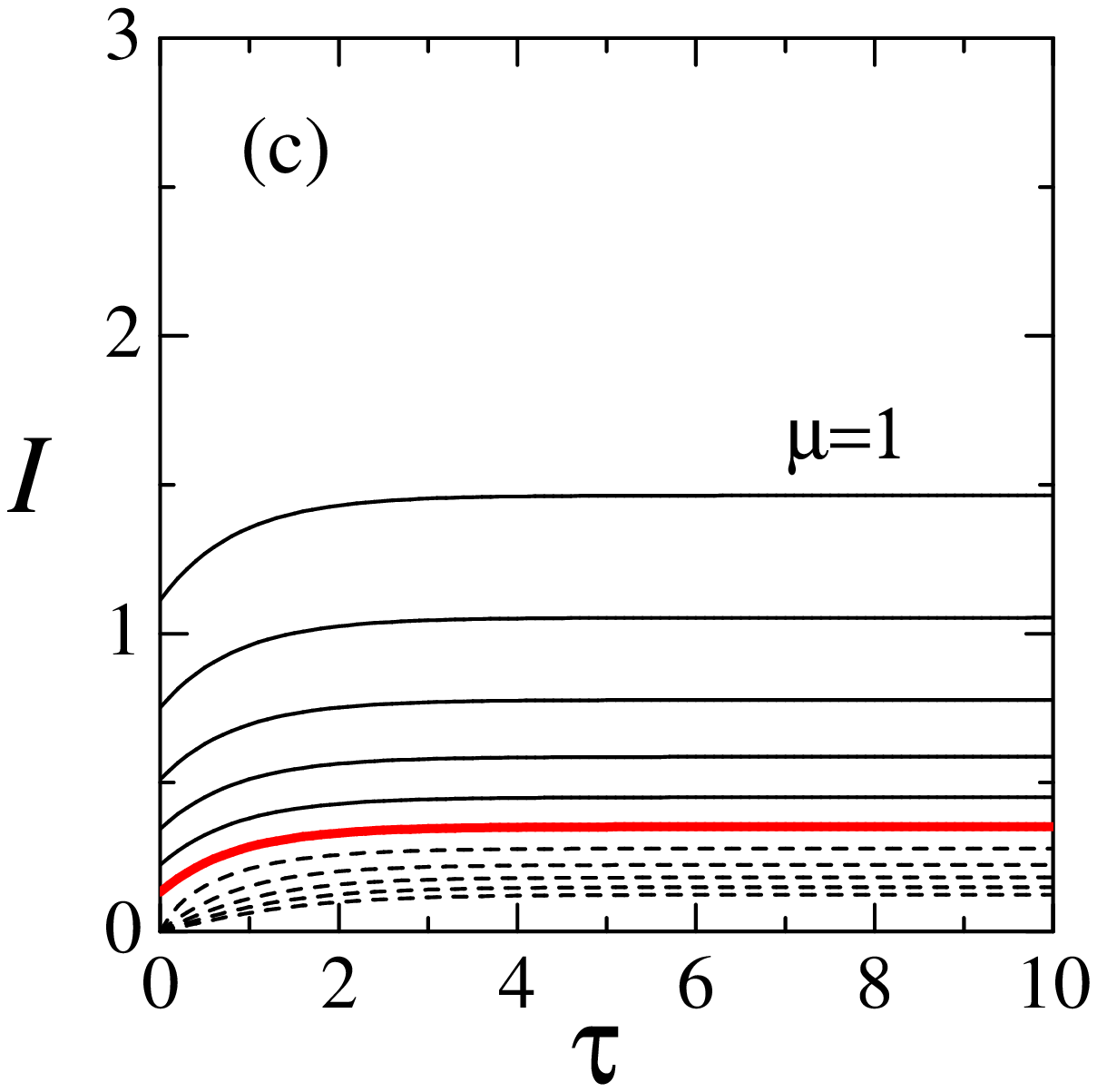}
  \end{center}
\caption{
Normalized specific intensity of Milne-Eddington solutions
for relativistic plane-parallel flows
as a function of the optical depth $\tau$
for several values of $\mu$ and $\beta$.
The values of $\beta$
are (a) 0, (b) 0.1, and (c) 0.3.
The values of $\mu$ are
0 (thick solid curves), $0.2, 0.4, 0.6, 0.8$, and 1 (solid curves),
and $-0.2, -0.4, -0.6, -0.8$, and $-1$ (dashed ones).
}
\end{figure}

In figure 2,
the specific intensity normalized by $I_{\rm s}$ of 
Milne-Eddington solutions of transfer equation
for relativistic plane-parallel flows
are shown as a function of the optical depth $\tau$
for several values of $\mu$ and $\beta$.
The values of $\beta$
are (a) 0, (b) 0.1, and (c) 0.3.
In each figure,
the thick solid curves are for $\mu=0$,
the solid curves are for $\mu=0.2, 0.4, 0.6, 0.8$,
and 1 from bottom to top, while
the dashed ones are for $\mu=-0.2, -0.4, -0.6, -0.8$,
and $-1$ from top to bottom.

As is seen in figure 2,
in the case of small $\beta$ (figure 2a),
the usual Milne-Eddington solution for a static plane-parallel
atmosphere is well reproduced.
As the flow speed increases,
the specific intensity becomes constant,
and is enhanced ($\mu>0$) or diminished ($\mu<0$)
due to the relativistic vertical motion.

In figure 3,
the emergent intensity $I(0, \mu, \beta)$
normalized by $I_{\rm s}$ of 
Milne-Eddington solutions of transfer equation
for relativistic plane-parallel flows
is plotted as a function of $\mu$ 
for several values of $\beta$.
The values of $\beta$
are 0, 0.1, 0.2, 0.3, 0.4, and 0.5.
The dashed line is for the usual
Milne-Eddinton solution for the static plane-parallel case.

In the limit of $\beta=0$,
the present emergent intensity again reduces to
the usual Milne-Eddington solution for a static atmosphere.
As the flow speed increases,
the {\it relativistic peaking effect} becomes prominent;
that is to say,
the emergent intensity is strongly enhanced
toward the poleward direction,
while it is less enhanced in the edgeward direction.

\begin{figure}
  \begin{center}
  \FigureFile(80mm,80mm){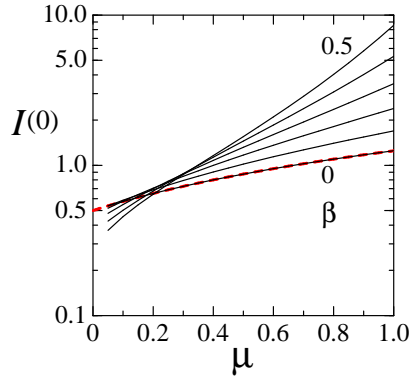}
  \end{center}
\caption{
Normalized emergent intensity of Milne-Eddington solutions
for relativistic plane-parallel flows
as a function of $\mu$
for several values of $\beta$.
The values of $\beta$
are 0, 0.1, 0.2, 0.3, 0.4, and 0.5.
The dashed line is for the usual
Milne-Eddinton solution for the static plane-parallel case.
}
\end{figure}

\section{Discussion}

In the previous sections
we have obtained the analytical solutions (\ref{E_sol})--(\ref{P_sol})
of relativistic moment equations,
and the analytical solutions (\ref{i_plus}) and (\ref{i_minus})
of relativistic transfer equation,
using the solutions (\ref{E_sol})--(\ref{P_sol}).
However, there is no assurance that
these solutions are consistent each other.
Hence, we here briefly discuss the consistency of the present solutions,
and check the accuracy of the present results.

Using the radiation fields $E$, $F$, and $P$ in the inertial frame,
we obtained the radiation intensity $I$ in the inertial frame.
Once the radiation intensity $I(\tau, \mu, \beta)$ is obtained,
we can calculate the radiation energy density 
$E_{\rm num}$ ($=2\pi \int I d\mu/c$),
the radiative flux $F_{\rm num}$  ($=2\pi \int I\mu d\mu$),
and the radiation pressure $P_{\rm num}$  ($=2\pi \int I\mu^2 d\mu/c$)
by the definition.
If the quantities obtained from the moment equations
and those obtained from the intensity coincide each other,
the solutions are consistent, and vice versa.

\begin{figure}
  \begin{center}
  \FigureFile(80mm,80mm){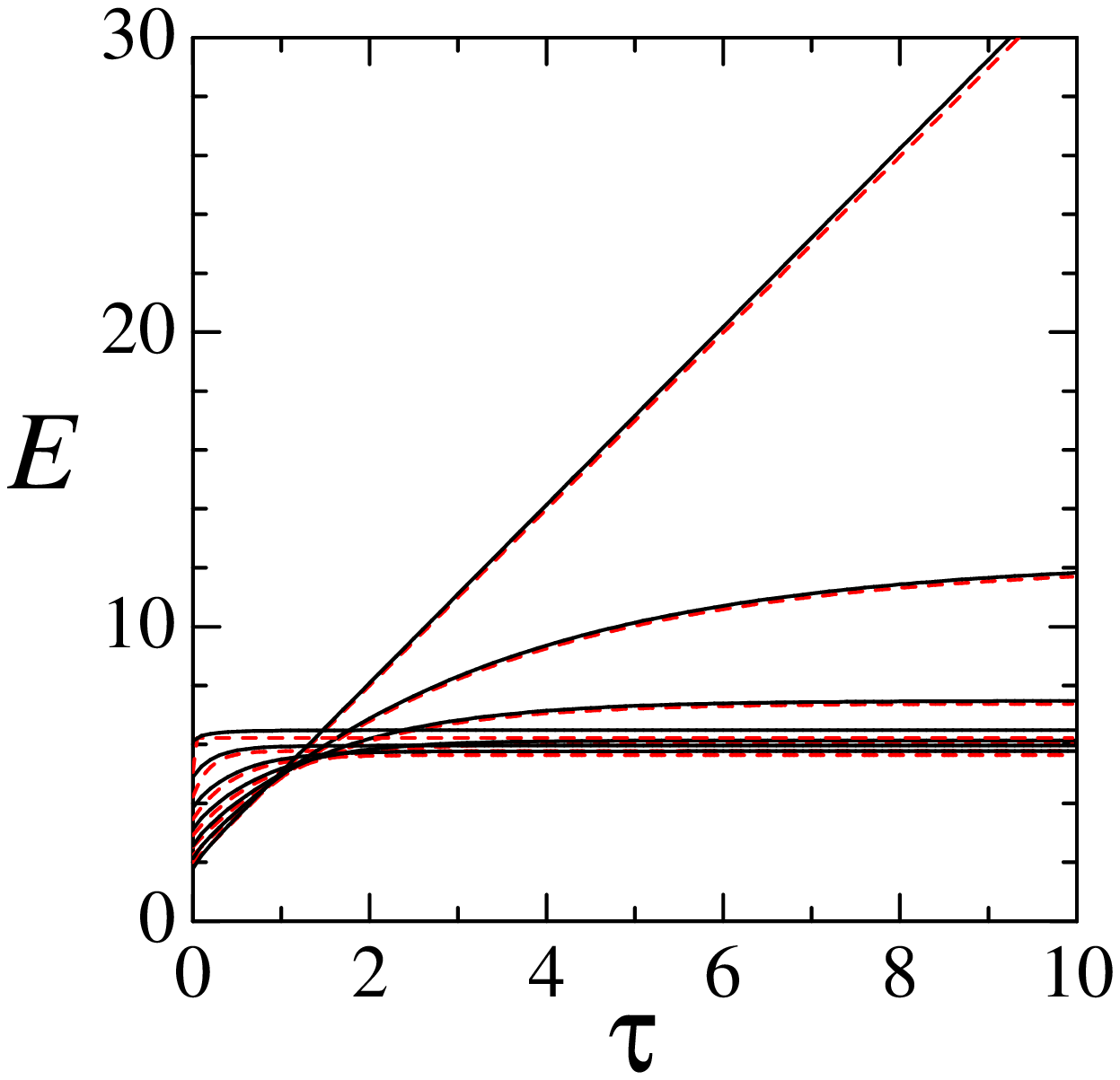}
  \FigureFile(80mm,80mm){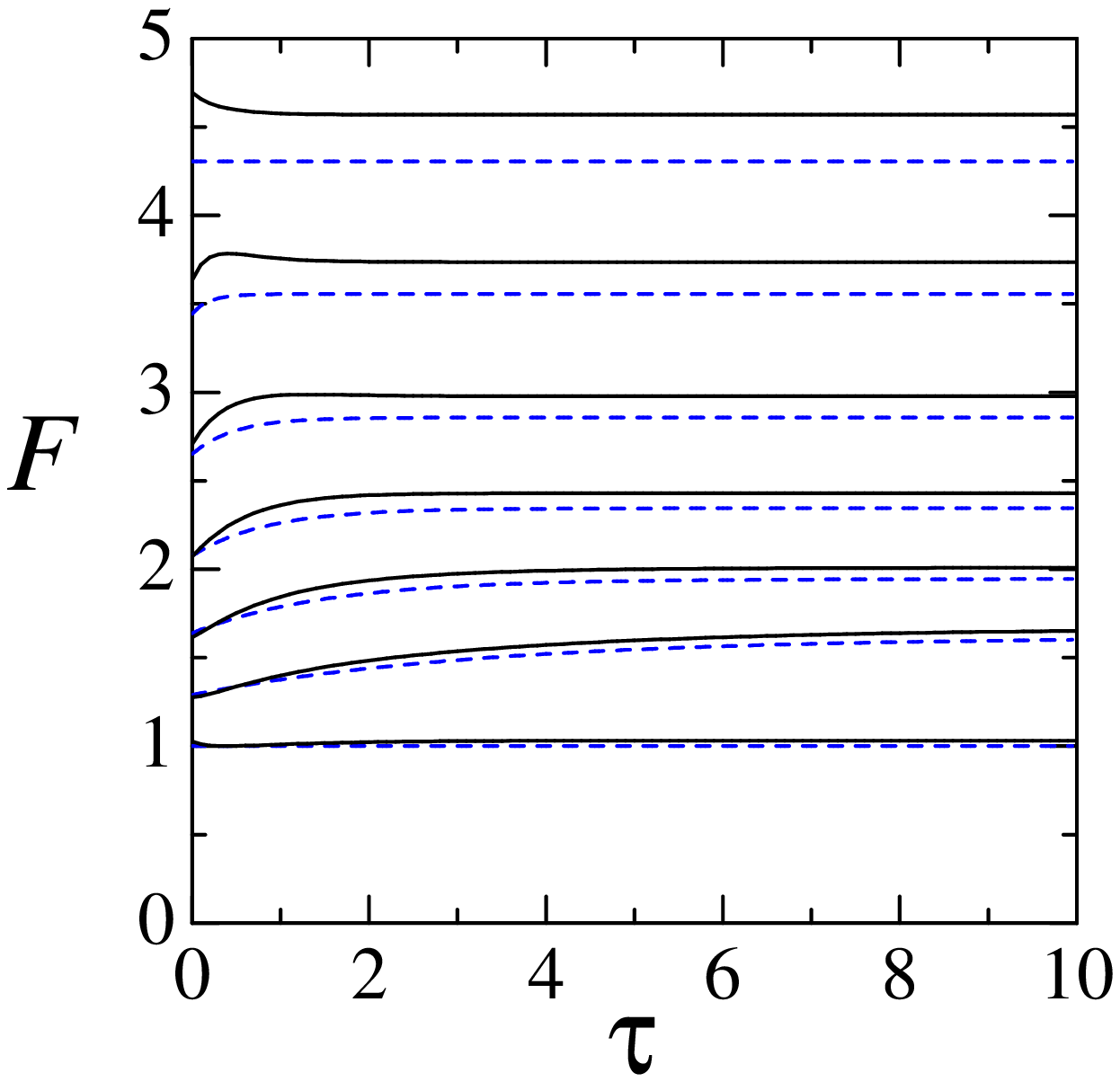}
  \FigureFile(80mm,80mm){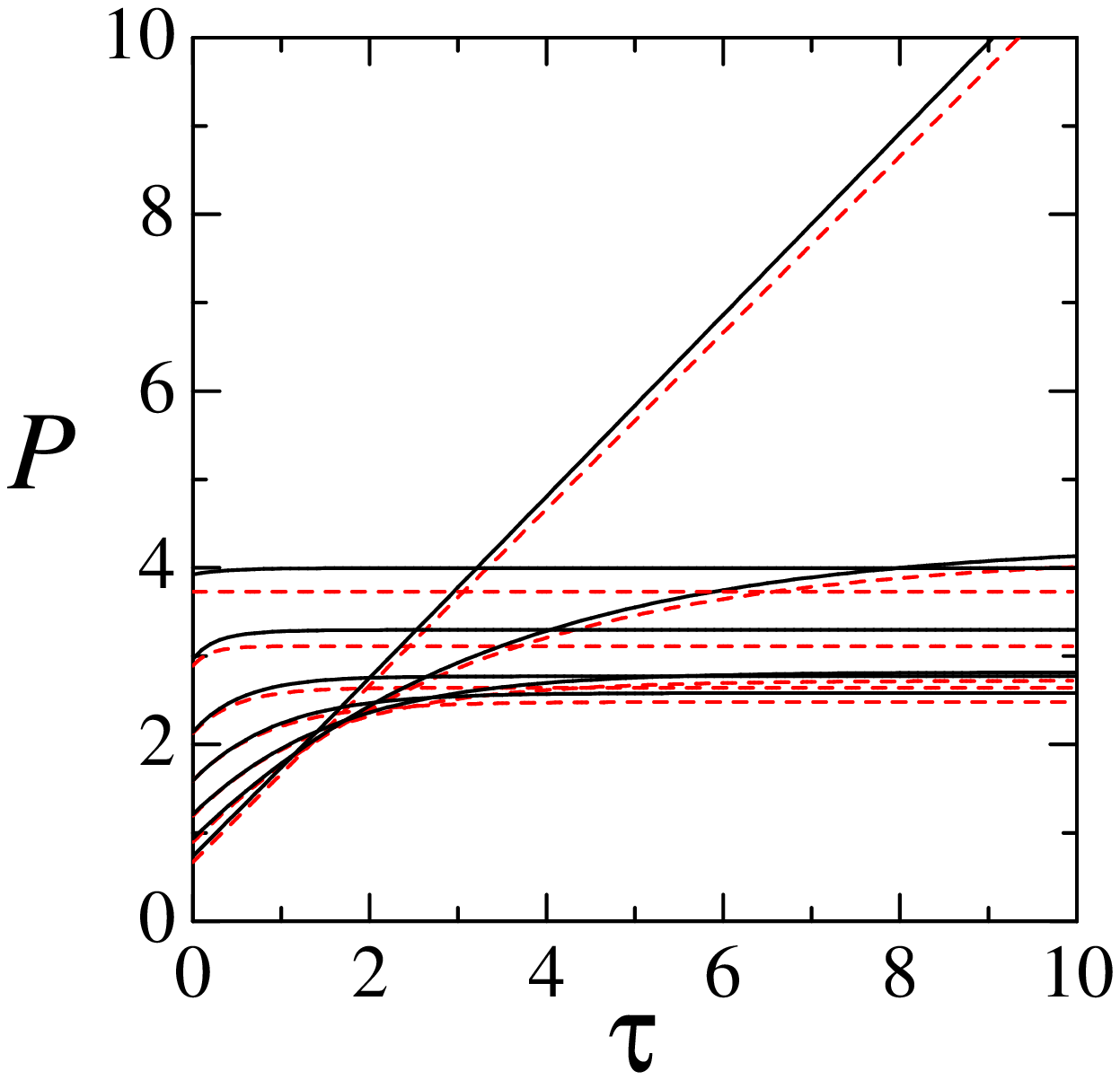}
  \end{center}
\caption{
Radiative quantities obtained from moment equations
and those obtained from the radiative intensity:
(a) Radiation energy density,
(b) radiative flux, and
(c) radiation pressure.
The dashed curves are the analytical solutions of moment equations and
the solid curves are the numerical solutions obtained by
the integration of the analytical solutions of transfer equation.
The values of $\beta$ are
0, 0.1, 0.2, 0.3, 0.4, 0.5, and 0.577.
}
\end{figure}

In figure 4
we show
the analytical solutions of moment equations (dashed curves)
and the numerical solutions obtained by the integration
of the analytical solutions of transfer equation (solid curves)
for various values of $\beta$:
(a) Radiation energy density $E$,
(b) radiative flux $F$, and
(c) radiation pressure $P$,
where the radiation quantities are normalized by $\pi I_{\rm s}$.
The flow speed $\beta$ are set to be
0, 0.1, 0.2, 0.3, 0.4, 0.5, and 0.577.

As is seen from figure 4,
in the case of small velocity,
both solutions are almost coincident.
As the flow speed increases,
the numerical solutions obtained from the intensity
become a little bit larger than the analytical solutions.
However, the discrepancy is not so large,
and within ten parcents at most.

\begin{figure}
  \begin{center}
  \FigureFile(80mm,80mm){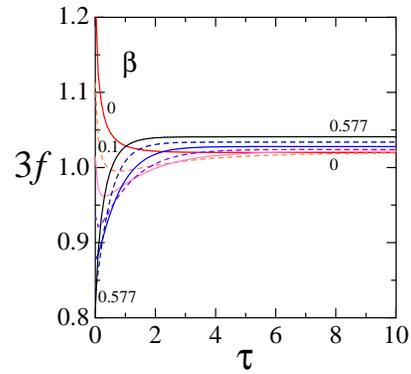}
  \end{center}
\caption{
Eddington factor obtained from the radiative intensity
for various values of the flow speed.
The values of $\beta$ are
0, 0.1, 0.2, 0.3, 0.4, 0.5, and 0.577.
}
\end{figure}

We can also calculate the Eddington factor $f$
by the numerical integration 
of the analytical solution $I(\tau, \mu, \beta)$.
In figure 5
the values of $3f$ are shown
for various values of $\beta$
($=$ 0, 0.1, 0.2, 0.3, 0.4, 0.5, and 0.577).

In the limit of $\beta=0$, i.e., in the case of the static atmosphere,
the Eddington factor slightly becomes larger than 1/3
for the small optical depth.
This is just the limb-darkening effect (or the peaking effect)
originated from thea anisotropic properties
of the radiation fields near to the surface.
Indeed, it is easy to derive that
the Eddington factor of the traditional static atmosphere
becomes $(17/14)(1/3)$ at $\tau=0$ (see, e.g., Fukue, Akizuki 2006a).

As the flow speed increases, however,
the Eddington factor near to the surface becomes 1/3,
and then further decreases to be smaller than 1/3.
This can be interpreted by the relativistic effect;
the relativistic aberration would cancell or dominate
the limb-darkening effect
in the region of small optical depths.
Anyway, except for the regions of $\tau < 1$,
the Eddington factor is almost 1/3
even if the flow speed is large.

%%%%%%%%%%%%%  CONCLUDING REMARKS  %%%%%%%%%%%%%%%%%%%%%%%%%%%%

\section{Concluding Remarks}

In this paper 
we have examined the radiative transfer problem
in the relativistic plane-parallel flows,
e.g., accretion disk winds,
under the assumption of a constant flow speed
in the fully special relativistic regime.
We have found the analytical solutions (\ref{E_sol})--(\ref{P_sol})
of the relativistic moment equations,
and the analytical solutions (\ref{i_plus}) and (\ref{i_minus})
of the relativistic transfer equation,
using solutions (\ref{E_sol})--(\ref{P_sol}).
These analytical solutions are
the {\it generalized Milne-Eddington solutions}
for the relativistically-moving plane-parallel flows.

In the subrelativistic regimes
many researchers have examined the radiative transfer
and radiation hydrodyanmics, and
the problem of radiation hydrodynamics
were often numerically solved
(e.g., Ohsuga et al. 2005; Ohsuga 2006;
see also Mihalas, Mihalas 1984).
It seems, however, that the radiative transfer problem
in the fully special relativistic regime
is not yet well understood.
Actually,
we found new analytical solutions,
and there would exist another solutions.
The analytical approach in the present work
 as well as the numerical approach
would help us to understand the relativistic radiative transfer.

In the present study
we assumed that the total optical depth is sufficiently large
and the flow speed is constant.
If the total optical depth is finite,
and the boundary condition at the flow base is taken account of,
the radiation fields and the resultant Eddington factor
would be changed
(cf. Fukue, Akizuki 2006a, b; Koizumi, Umemura 2007).
Or, if the velocity is not constant, but
there is a velocity gradient,
then the Eddington factor is proved to be less than 1/3
deep inside the flow (Fukue 2008b).
Such cases should be examined in the future.

We found that the emergent intensity depends
on the flow speed as well as the direction cosine,
and exhibits a relativistic peaking effect.
As a result,
a wind luminosity would be overestimated by a pole-on observer
and underestimated by an edge-on observer,
when we observe an accretion disk wind
(cf. Sumitomo et al. 2007).

We have also assumed the frequency-independent opacities
(gray approximation).
This is an approximation,
and can be extended using mean opacities
obtained from a frequency-dependent radiative transfer solution
(cf. Mihalas, Auer 2001).
In addition, we have obtained the frequency-integrated intensity $I$,
whereas the frequency-dependent intensity $I_\nu$
should be also examined in the future.

Furthermore, we have imposed various assumptions,
including cold approximation,
no heating source, and so on.
By relaxing these assumptions
and integrating the relativistic transfer equation numerically,
we could obtain the emergent intensity and spectra
more quantitatively.
These are also left as future works.

\vspace*{1pc}

%The author would like to thank
%S. Kato, S. Mineshige, M. Umemura, T. Koizumi, and C. Akizuki
%for enlightening and stimulating discussions.
%The authors would like to thank Professor Y. Osaki and Dr. N. Shibazaki
%for their valuable comments and discussions.
%The author would like to thank an anonymous referee for valuable comments.
This work has been supported in part
by a Grant-in-Aid for Scientific Research (18540240 J.F.) 
of the Ministry of Education, Culture, Sports, Science and Technology.

%\noindent

\end{document}